\documentclass[runningheads]{llncs}
\usepackage{graphicx}
\usepackage{subcaption}
\usepackage{float}
\usepackage{orcidlink}
\hypersetup{hidelinks}
\usepackage{multirow}
\usepackage{multicol}
\usepackage{pgfplots}
\usepackage{adjustbox}
\pgfplotsset{compat=1.17}
\usepackage[linesnumbered,ruled,vlined]{algorithm2e}

\usepackage{xcolor}
\newcommand\js[1]{{#1}} 
\newcommand\yh[1]{{#1}} 
\newcommand\jh[1]{{#1}} 
\newcommand\hw[1]{{#1}} 

\begin{document}
\title{MPSeg : Multi-Phase strategy for coronary artery Segmentation}
%
\author{
Jonghoe Ku\inst{1}\orcidlink{0009-0003-1260-619X} \and
Yong-Hee Lee\inst{1}\orcidlink{0000-0001-6047-701X} \and
Junsup Shin\inst{1}\orcidlink{0000-0003-3280-1622} \and
In Kyu Lee\inst{1}\orcidlink{0000-0001-5554-808X} \and
Hyun-Woo Kim\inst{1, *}\orcidlink{0009-0003-2740-0397}
}
\authorrunning{J.H. Ku et al.}
%
\institute{Medipixel Inc, Seoul, Republic of Korea \\
\email{\inst{*}hyunwoo.kim@medipixel.io}\\
}
\maketitle              

\begin{abstract}
\hw{
Accurate segmentation of coronary arteries is a pivotal process in assessing cardiovascular diseases. However, the intricate structure of the cardiovascular system presents significant challenges for automatic segmentation, especially when utilizing methodologies like the SYNTAX Score, which relies extensively on detailed structural information for precise risk stratification. To address these difficulties and cater to this need, we present MPSeg, an innovative multi-phase strategy designed for coronary artery segmentation. Our approach specifically accommodates these structural complexities and adheres to the principles of the SYNTAX Score. Initially, our method segregates vessels into two categories based on their unique morphological characteristics: Left Coronary Artery (LCA) and Right Coronary Artery (RCA). Specialized ensemble models are then deployed for each category to execute the challenging segmentation task. Due to LCA's higher complexity over RCA, a refinement model is utilized to scrutinize and correct initial class predictions on segmented areas. Notably, our approach demonstrated exceptional effectiveness when evaluated in the Automatic Region-based Coronary Artery Disease diagnostics using x-ray angiography imagEs (ARCADE) Segmentation Detection Algorithm challenge at MICCAI 2023.}
\end{abstract}
\section{Introduction}

\yh{

Coronary artery disease (CAD) is the leading cause of mortality worldwide~\cite{dalen14,cad_lead_to_16,roth17,who_cvd_rpt}.
The diagnosis and treatment of CAD is important within the medical domain.
CAD often exhibits intricate and diverse physiological aspects, including anatomical variance, multi-vessel diseases, and complex lesions, which should be considered for optimal treatment~\cite{complex_02,aha_guide_23}.

There have been some attempts to systemically evaluate CAD due to its high complexity.
In 1975, the American Heart Association suggested the reporting system for evaluating CAD~\cite{aha_def_seg_75}.
In the reporting system, the coronary tree segments are defined, which is modified later for the ARTS (Arterial Revascularization Therapies Study) I and II trials~\cite{art_syn_first_07,artsII_5y_10}.
The modified definition is utilized for calculating the SYNTAX (SYNergy between PCI with TAXUS\textsuperscript{TM} and Cardiac Surgery) Score~\cite{syntax_origin}.
The SYNTAX Score is an evaluation system developed to assist the risk stratification of patients by integrating the validated angiographic classifications, including the morphology and location of CAD.

The structure of the coronary artery is complex and personalized. 
In general, the coronary artery consists of two main branches, LCA and RCA.
LCA is divided into left anterior descending (LAD) and circumflex artery (LCx) after left main, then side branches extend from the main vessels, LAD, LCx, and RCA, covering the epicardial surface.
In the real world, the clinical operator confronts various vasculature from normal variants, such as right or left dominance, to anomalies, such as separated origins of the LAD and LCx~\cite{anomalies_16}.
The heterogeneity and the complexity of the coronary vessel structure hinder the inexperienced operator from performing robust segmentation of vessel segments~\cite{syntax_origin}.

Moreover, manual segmentation of vessel segments is not only time-consuming but also susceptible to intra- and inter-operator variability. These factors can lead to inconsistent analysis results and pose challenges in achieving reliability and accuracy. Therefore, there exists a compelling need for the development of automatic segmentation approaches that can provide both efficiency and precision in the segmentation of vessel structures. Such automated methods have the potential to significantly enhance the speed and consistency of medical image analysis, ultimately benefiting both patients and clinical operators.
}

\begin{figure}[t]
\includegraphics[width=\textwidth]{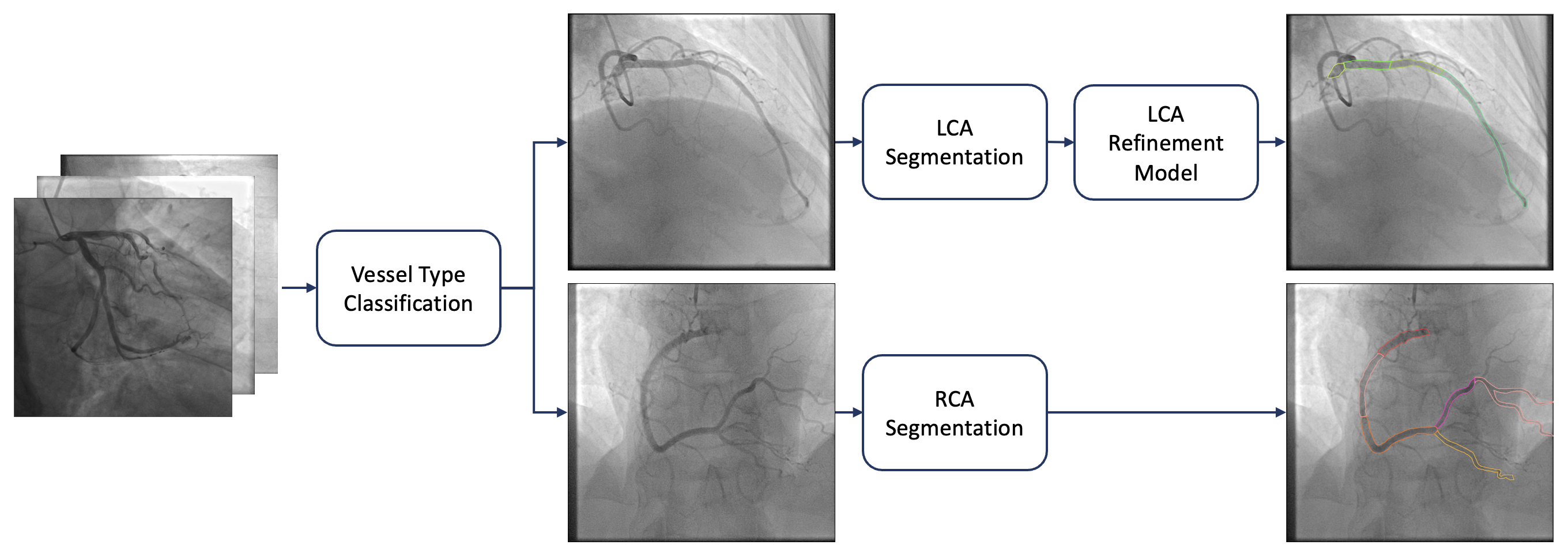}
\caption{Visualization of the entire pipeline in the proposed methodology.}
\label{fig:overall_pipeline}
\end{figure}

\js{
In recent years, the rapid advancements in machine learning have significantly impacted various fields, notably in the domain of medical image processing~\cite{ref_medical_img_seg}. Segmentation in medical image processing has garnered substantial attention in prior studies~\cite{ref_unet,ref_medical_img_seg,ref_breast_tumor}.
UNet~\cite{ref_unet}, a well-known model in the domain, is renowned for its symmetric structure and utilization of skip connections, enabling effective learning of low and high-level features.
Cho et al.~\cite{ref_breast_tumor} presented a multi-stage approach for segmenting breast tumors.
The authors initially trained a binary classification model to detect the presence of a tumor and subsequently focused on segmenting the identified region.
Also, the authors contended that this two-stage strategy significantly reduces false positives, and this strategy aligns with our approach.



}

\js{
We conducted a thorough investigation of the coronary artery segmentation dataset, specifically focusing on the challenges related to the SYNTAX Score segment.
This analysis guided us in streamlining the intricacies of the task, resulting in a more manageable approach.
Therefore, the proposed methodology strategically addresses the complexities through a well-defined multi-stage strategy.
}

\js{
In the coronary tree segmentation challenge, the proposed method showcased three distinctive features:
\begin{itemize}
    \item \textbf{Data Analysis}: Leveraging extensive exploratory data analysis on the ARCADE 2023 challenge dataset, we meticulously examined the intricacies of coronary artery segmentation in the context of the SYNTAX Score segment. This analysis was pivotal in simplifying the challenges associated with the task.
    \item \textbf{Multi-stage Strategy}: Our proposed methodology is intentionally crafted to navigate and resolve the intricacies of the complication using a thoughtfully devised multi-stage strategy. This strategic approach contributes to the efficiency and efficacy of our solution.
    \item  \textbf{Exceptional Efficacy}: The proposed method demonstrated outstanding performance in the ARCADE segmentation task in the competition, underscoring its exceptional efficacy.
\end{itemize}
}

\js{

The proposed method encompasses three pivotal stages: Vessel classification, SYNTAX segmentation, SYNTAX classification.
In the initial vessel classification stage, the image is subject to binary classification, distinguishing between RCA and LCA. Since the images manifest distinct characteristics between RCA and LCA but not within LCA (specifically LCx and LAD), the proposed method first classifies the vessels of interest as either belonging to RCA or LCA.
After the classification, the vessels are segmented based on the SYNTAX Score segments. Two separate models are employed for classification output, one for RCA and another for LCA. In the case of classified as RCA, the predicted SYNTAX Score segment directly serves as the output, given the assumption that the image of RCA is relatively less complex than LCA views. On the other hand, for the images classified as LCA, the last stage SYNTAX classification is used to classify the final prediction of the segment by utilizing the original image and the mask derived from the segmentation stage. This classification is performed specifically within the masked region, determining the SYNTAX Score segment category accurately.
Detailed methodologies for each stage are elaborated in section 2.

}

\section{Method}

\jh{
In this section, we present an overview of the ARCADE challenge dataset, provide an analysis of the segmentation dataset, and describe the evaluation metrics utilized in the competition. Furthermore, we elaborate on the approaches employed for SYNTAX segmentation, including the models used. Subsequently, we delve into details related to training.
}

\jh{
\subsection{ARCADE Challenge Dataset}
\subsubsection{Dataset}
The ARCADE challenge's coronary tree segmentation dataset consists of 1000 training images, 200 validation images, and 300 test images. This dataset has 25 SYNTAX Score classes, with 8 classes assigned to RCA and 17 classes allocated to LCA.
}
\jh{
\begin{figure}[t]
    \centering
    \begin{adjustbox}{width=1.0\textwidth, center}
    \begin{tikzpicture}
        \begin{axis}[
            ybar,
            width=1.0\textwidth,
            height=0.4\textwidth,
            bar width=10pt,
            ylabel={Number of segments},
            xlabel={SYNTAX Score class},
            symbolic x coords={1,2,3,4,5,6,7,8,9,9a,10,10a,11,12,12a,12b,13,14,14a,14b,15,16,16a,16b,16c},
            xtick=data,
            x tick label style={rotate=45,anchor=east},
            nodes near coords={\scriptsize\pgfmathprintnumber\pgfplotspointmeta},
            nodes near coords align={vertical},
            enlarge x limits={abs=0.5cm}
            ]
            \addplot coordinates {
                (1,374) (2,375) (3,369) (4,303) (5,527) (6,536) (7,340) (8,310) (9,198) (9a,70) (10,21) (10a,1) (11,319)
                (12,61) (12a,129) (12b,305) (13,107) (14,49) (14a,38) (14b,231) (15,43) (16,48) (16a,31) (16b,63) (16c,127)
            };
        \end{axis}
    \end{tikzpicture}
    \end{adjustbox} 
    \caption{\label{fig:eda_histogram}Distribution of SYNTAX Score classes in ARCADE challenge vessel tree segmentation task}
\end{figure}
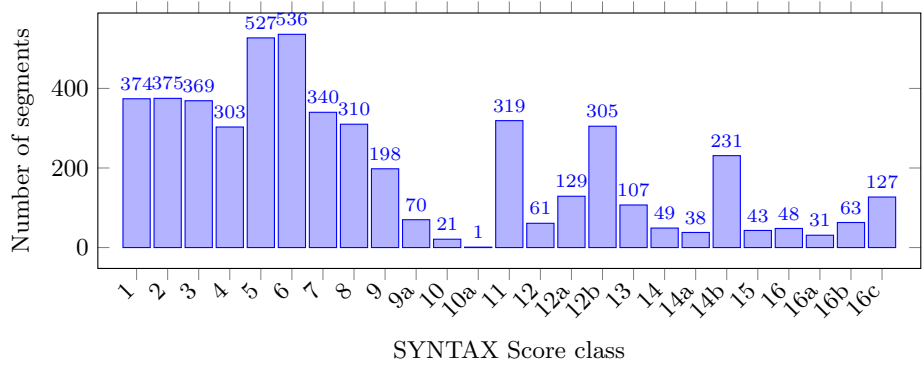

\subsubsection{Exploratory Data Analysis}

When we analyzed the segmentation dataset, the ratio of RCA to LCA vessel types was approximately 1:2. Furthermore, when examining the distribution of SYNTAX segment classes, a significant imbalance was identified between the main arteries and side branches. In Fig.~\ref{fig:eda_histogram}, there were instances where the number of side branch segments was substantially small, and certain classes had only one case present in the training data.  
To address this imbalance, we aimed to train separate segmentation models for each type of vessel. Vessel classification was carried out as described in Algorithm \ref{algo:coronary_classification}. While this approach effectively classified RCA and LCA, the classification of LCx and LAD proved challenging due to the presence of mixed data from both vessels, making accurate classification unfeasible. Consequently, the final choice for vessel types to be used in the classification was narrowed down to RCA and LCA.
}
\begin{algorithm}[h]
\caption{Coronary Artery Classification based on SYNTAX Score Index}\label{algo:coronary_classification}
\SetAlgoLined
\KwIn{List of SYNTAX Score index labels}
\KwOut{Coronary artery classification}
\BlankLine

\eIf{segment label contains 1, 2, 3, 4, 16, 16a, 16b, 16c}{
    Coronary artery classification $\leftarrow$ RCA\;
}{
    \eIf{segment label contains 11, 12, 13, 14, 14a, 14b, 15}{
        Coronary artery classification $\leftarrow$ LCX\;
    }{
        Coronary artery classification $\leftarrow$ LAD\;
    }
}
\end{algorithm}

\jh{
\subsubsection{Evaluation}

To evaluate SYNTAX segmentation performance, the challenge employed mean F1 score.

The F1 score is calculated by precision and recall.
\begin{equation}
F1 = 2 \frac{precision\cdot recall}{precision+recall}
\end{equation}
Precision and recall can be calculated using True Positive (TP), False Positive (FP) and False Negative (FN).
\begin{equation}
precision = \frac{TP}{TP+FP}
\end{equation}

\begin{equation}
recall = \frac{TP}{TP+FN}
\end{equation}
The F1 score is individually calculated for each class, considering only one segment at a time. When computing the F1 score, it is based on the classes present in the ground truth. If the model predicts a class that is not included in the ground truth, that prediction is disregarded in the evaluation process. To measure the mean F1 score, we aggregated all the individual F1 scores and divided this sum by the total number of segments for which F1 scores were computed.
\begin{equation}
mean F1 = \frac{1}{N} \sum_{i=1}^{N} \frac{1}{C_i} \sum_{j=1}^{C_i}  F1_{ij}
\end{equation}
$N$ and $C$ denote the total number of images and the number of segmented classes of each image, respectively.
}

\jh{
\subsection{Proposed Method}
%
}

\subsubsection{Data Augmentation}

\jh{
Accurate capturing of the SYNTAX segment necessitates an understanding of the spatial relationships within the entire vascular structure. To ensure the preservation of these relationships, we limit our data augmentations to basic transformations such as rotation and translation. This approach allows us to make minor adjustments while avoiding more drastic modifications like mosaic or flipping, which could potentially disrupt the delicate spatial information essential for accurate segmentation.

Furthermore, coronary images are acquired in grayscale with varying brightness distribution and noise characteristics depending on the imaging equipment and environmental conditions. To address these variations, we include augmentations such as adjusting brightness and blurring to ensure reliable performance across a range of imaging conditions encountered in practice.
}

\subsubsection{Vessel classification}
\jh{
The morphology of the RCA and LCA exhibits distinct differences. Therefore, it is better to train a classification model to distinguish between these two types of vessels instead of trying to predict the entire SYNTAX segment in one step during SYNTAX segmentation.

To achieve this segmentation based on the classification of vessels, a preliminary step is taken by employing a vessel classification model. 
This model takes a coronary image as an input and categorizes it into either the RCA or LCA, as shown in Fig.~\ref{fig:overall_pipeline}.
}

\subsubsection{SYNTAX segmentation}
\jh{
The input for SYNTAX segmentation is an image with dimensions $(512, 512, 1)$. Models trained on RCA images predict masks of size $(512, 512, 8)$, while models trained on LCA images predict masks of size $(512, 512, 17)$. This class separation strategy is implemented to simplify the training process and mitigate the risk of misclassification. It ensures that RCA SYNTAX segments are not predicted for LCA vessels and vice versa.

In the case of LCA, there are complexities in imaging angles and vessel structures that can lead to misclassification of side SYNTAX segments as shown in Fig.~\ref{fig:misidentification}. 
To overcome this issue, an additional classification model is used to adjust the class assignments for SYNTAX segments.

On the other hand, SYNTAX segments of RCA have clear distinctions between main and side segments, and the morphology of side segments is simpler than that of LCA's side segments.
This simplification in the classification process is facilitated by the inherent characteristics of the RCA, which makes it more amenable to direct segmentation without the need of additional classification refinement.

\begin{figure}
\includegraphics[width=6cm]{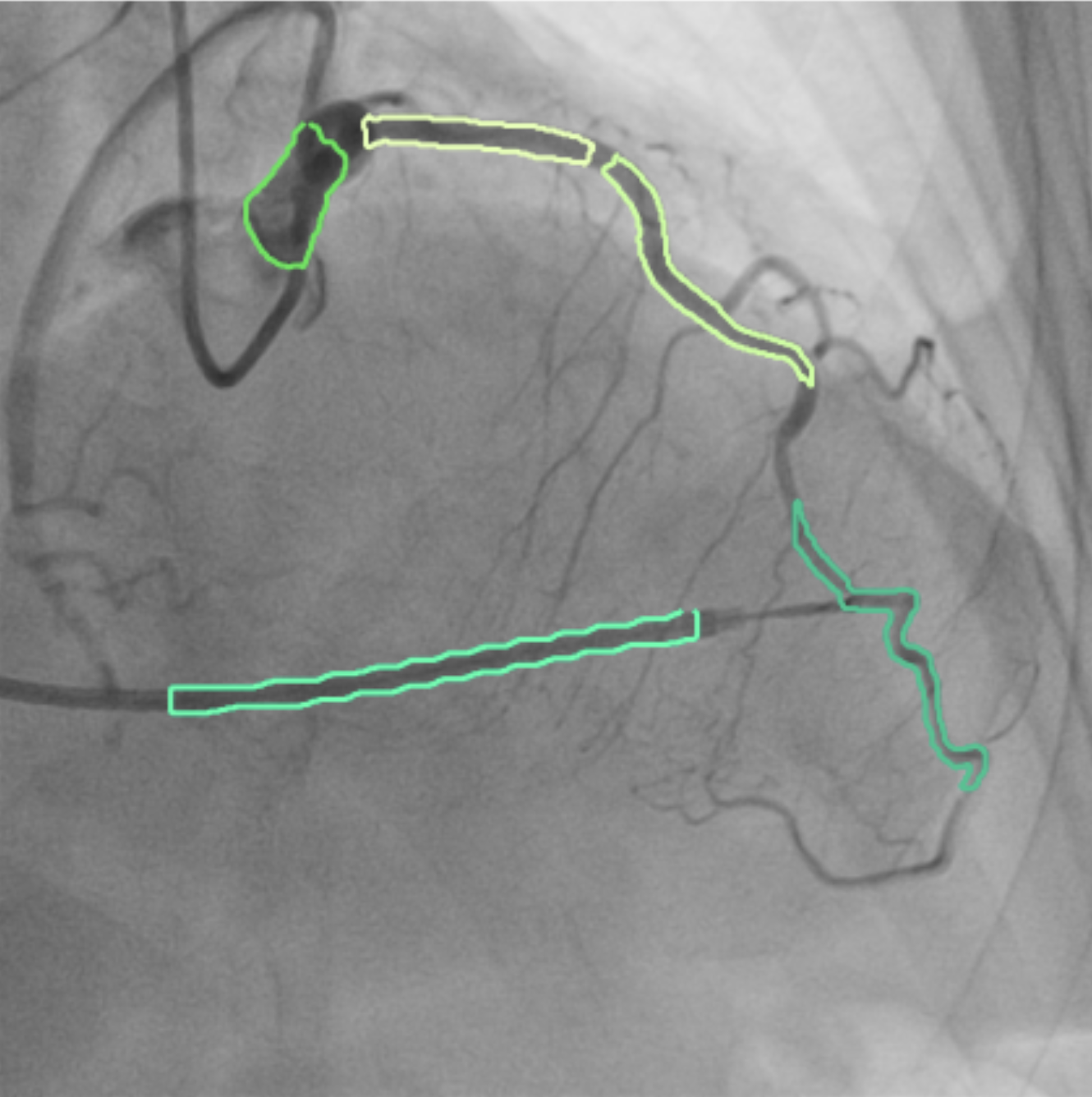}
\centering
\caption{An example of misidentified RCA segment in LCA. When a guide wire is located in a similar position to RCA, the segmentation model may recognize the guide wire as a segment of RCA.}
\label{fig:misidentification}
\end{figure}
}

\subsubsection{SYNTAX segment classification}
\jh{
We trained EfficientNet-b3\cite{ref_efficient}, ResNet34\cite{ref_resnset}, and DenseNet121\cite{ref_densenet} for SYNTAX segment classification. SYNTAX segment classification is exclusively conducted for LCA vessels, taking inputs in the form of coronary images and SYNTAX segment masks with dimensions $(512, 512, 2)$. Its purpose is to predict the LCA SYNTAX segment classes based on the image concatenated with the provided mask.

\begin{figure}
\includegraphics[width=\textwidth]{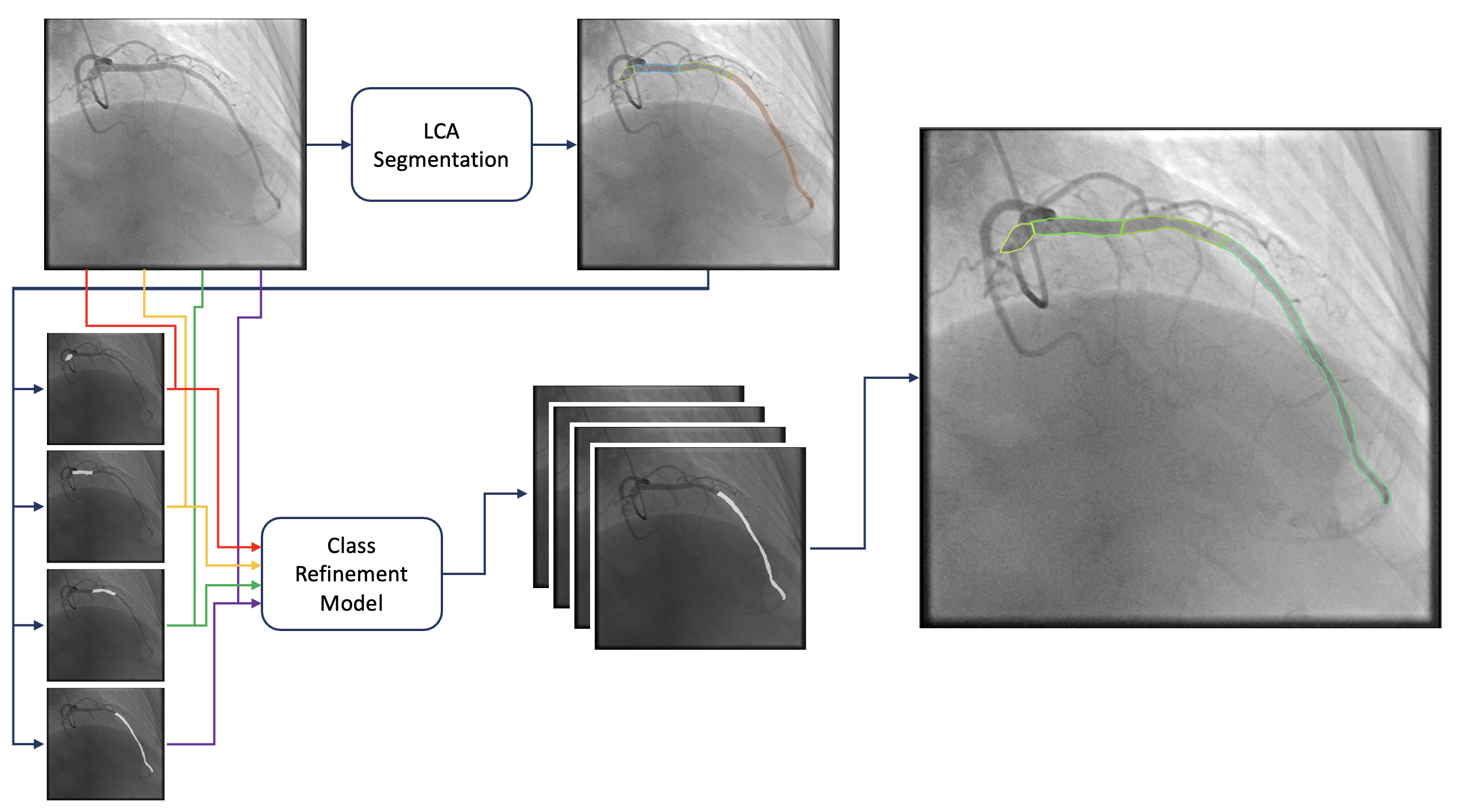}
\caption{Illustration of the internal workflow of the LCA refinement module in Fig~\ref{fig:overall_pipeline}.}
\label{fig:lca_refinement_pipeline}
\end{figure}
}

\subsection{Implementation details}
\jh{
The YOLOv8m model was trained for vessel classification using stochastic gradient descent (SGD) as the optimizer with a learning rate of 0.01. An ensemble approach was employed for the SYNTAX segmentation model, utilizing multiple models tailored to different vessel classes. For RCA, a UnetPlusPlus\cite{ref_unet++} model with a ResNet34 encoder was used, trained with 5-fold cross-validation using 800 training images and 200 validation images out of a total of 1000 training images. Among these, the top-performing three models were selected. Additionally, another UnetPlusPlus model with a ResNet50 encoder was used and trained on the entire set of 1000 training images.

For LCA, models were trained using a ResNet34 encoder with both UnetPlusPlus and Unet architectures, and a model with a ResNet50 encoder using UnetPlusPlus. Each of these models was trained using the AdamW\cite{ref_adamw} optimizer with a learning rate of 0.001 and focal loss\cite{ref_focal_loss}.

Finally, we performed an ensemble of these trained models to obtain the ultimate SYNTAX segmentation results on a vessel-by-vessel basis. For the SYNTAX segment classification in LCA, EfficientNet-b3, ResNet34, and DenseNet121 models were trained, and the final class was determined using an ensemble approach. The mean F1 score for the final segmentation is shown in Table \ref{total mean f1}.
}

\begin{table}[h]
\caption{Mean F1 score with and without SYNTAX classification.}\label{total mean f1}
\centering
\begin{tabular}{|c|c|c|}
\hline
Dataset & without SYNTAX classification  & with SYNTAX classification\\
\hline
Validation  & 0.482 &  \textbf{0.490}  \\
\hline
Test & 0.441 & \textbf{0.444}\\
\hline
\end{tabular}
\end{table}

\section{Result}
\begin{figure}[H]
\centering
\begin{tabular}{ccc}
\includegraphics[width = 1.5in]{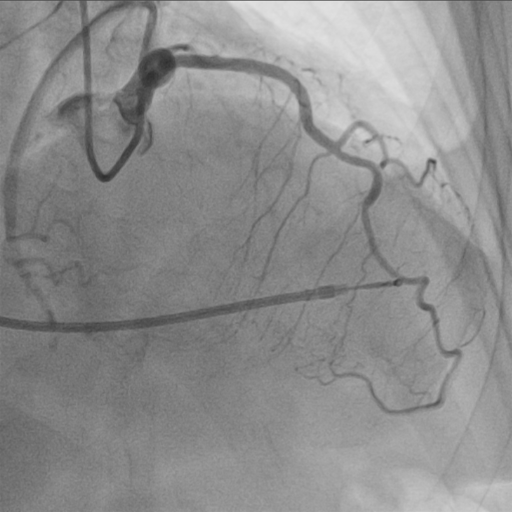} &
\includegraphics[width = 1.5in]{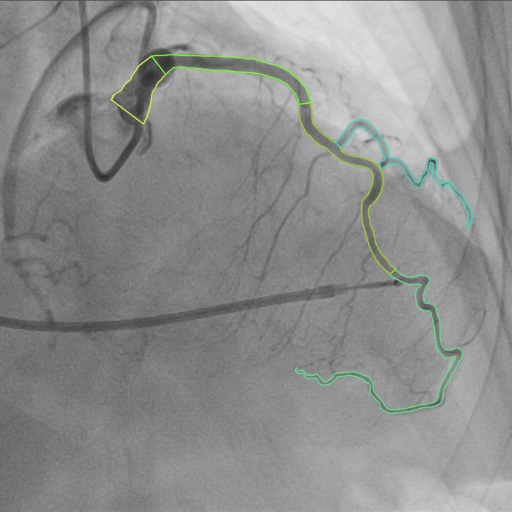} &
\includegraphics[width = 1.5in]{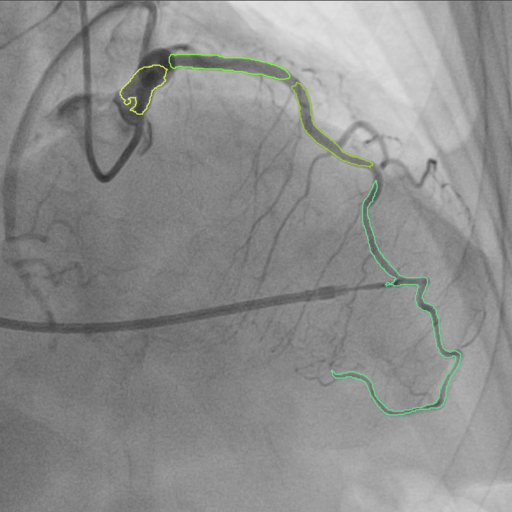} \\

\includegraphics[width = 1.5in]{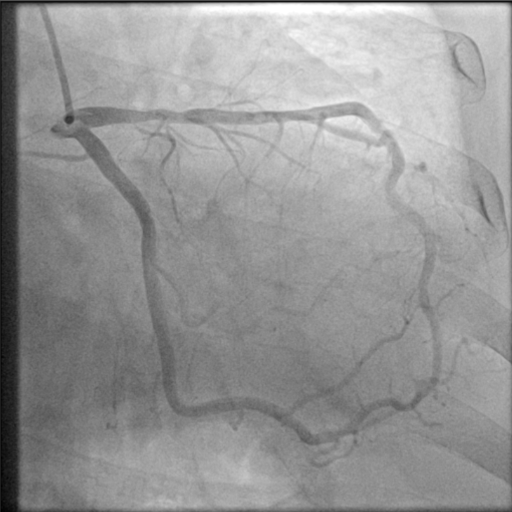} &
\includegraphics[width = 1.5in]{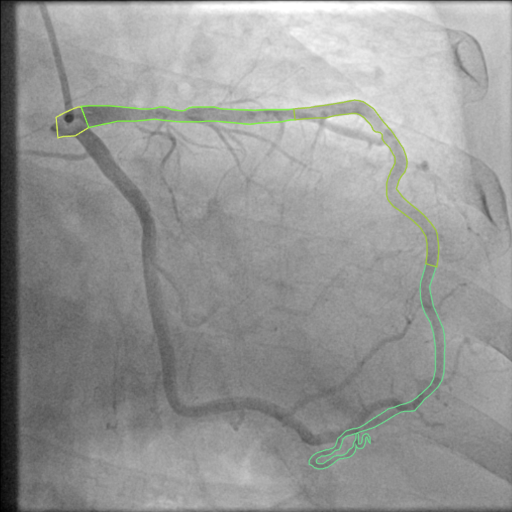} &
\includegraphics[width = 1.5in]{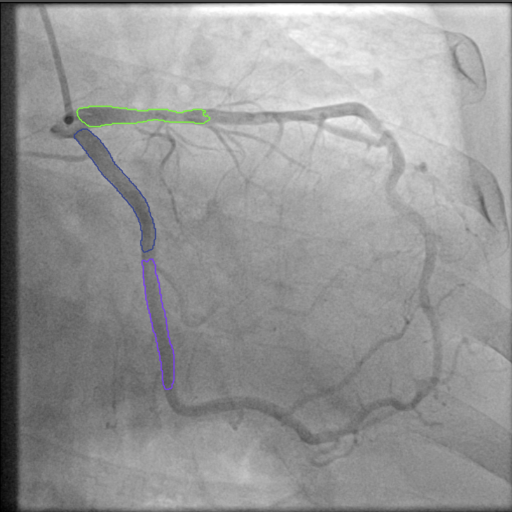} \\

\includegraphics[width = 1.5in]{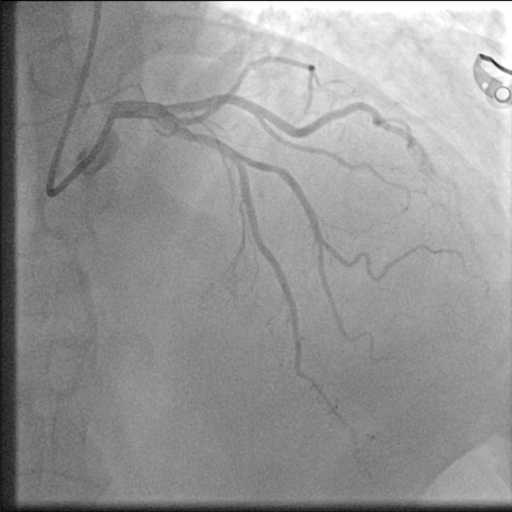} &
\includegraphics[width = 1.5in]{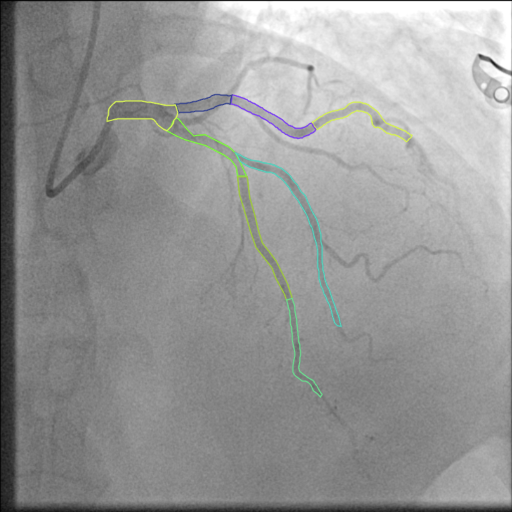} &
\includegraphics[width = 1.5in]{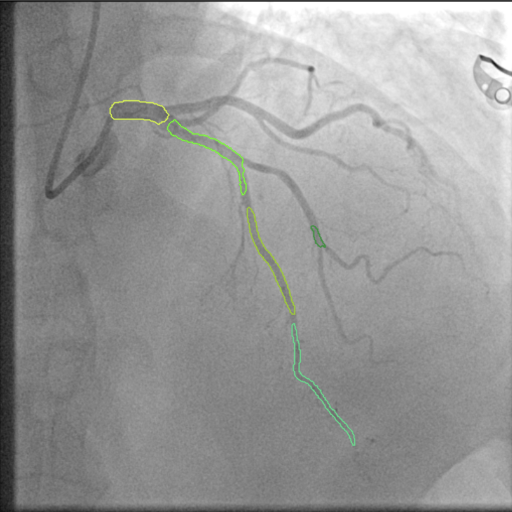} \\

\includegraphics[width = 1.5in]{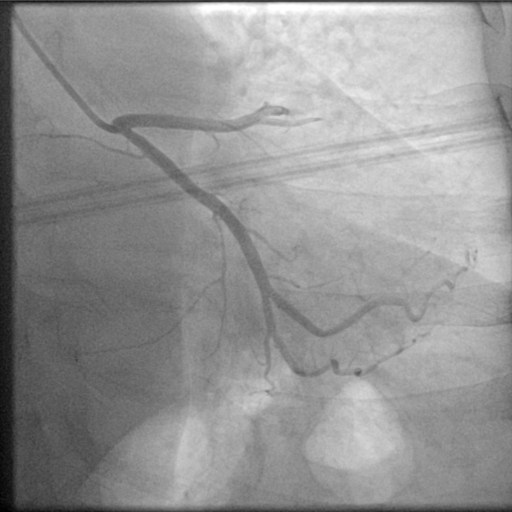} &
\includegraphics[width = 1.5in]{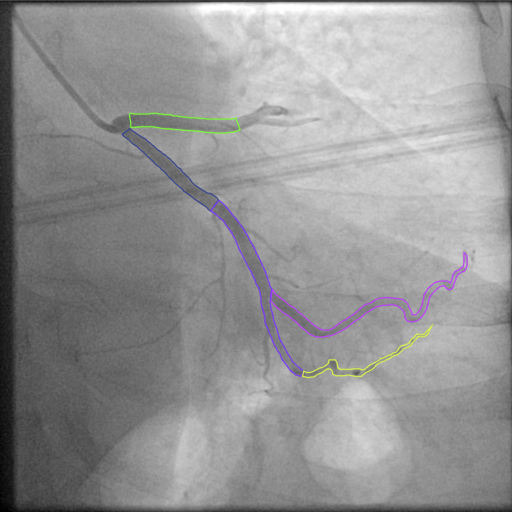} &
\includegraphics[width = 1.5in]{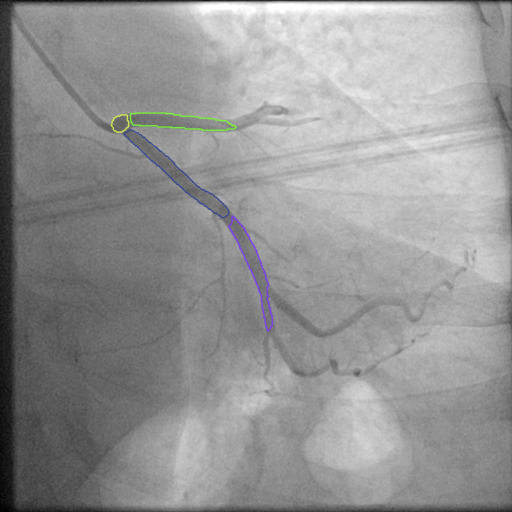} \\

\subfloat[Input]{\includegraphics[width = 1.5in]{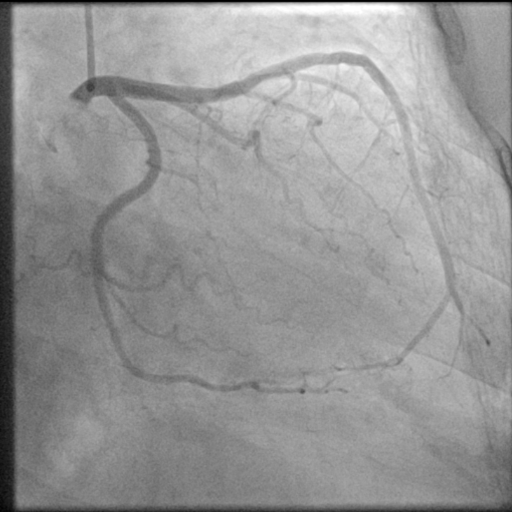}}&
\subfloat[Ground-truth]{\includegraphics[width = 1.5in]{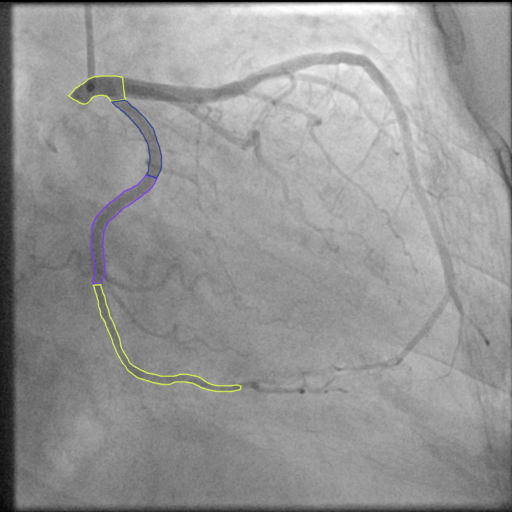}}&
\subfloat[Ours]{\includegraphics[width = 1.5in]{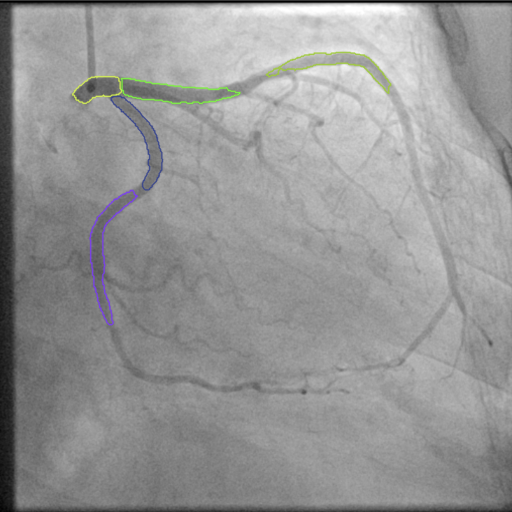}}
\end{tabular}
\caption{Qualitative results of LCA. Each SYNTAX segment class is distinguished by colors. The ensemble model predicts the main segment more precisely than the side segment. The coronary images in the 2nd and 5th rows exhibit a similar shape, but they have different ground truth classes.}
\label{fig:qualitative lca}
\end{figure}

\begin{figure}[H]
\centering
\begin{tabular}{ccc}
\includegraphics[width = 1.5in]{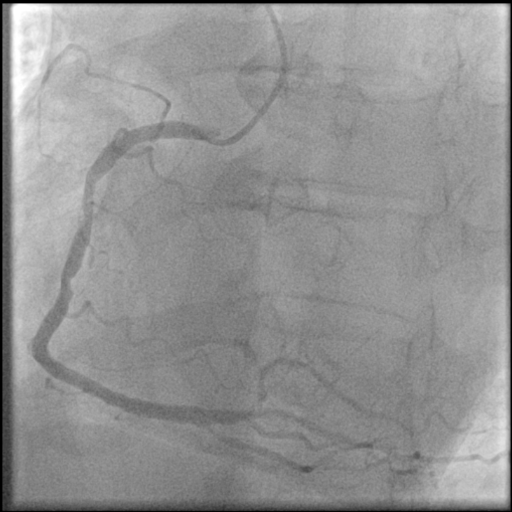} &
\includegraphics[width = 1.5in]{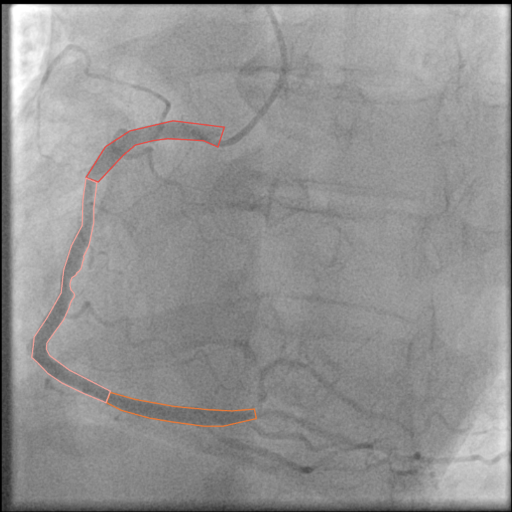} &
\includegraphics[width = 1.5in]{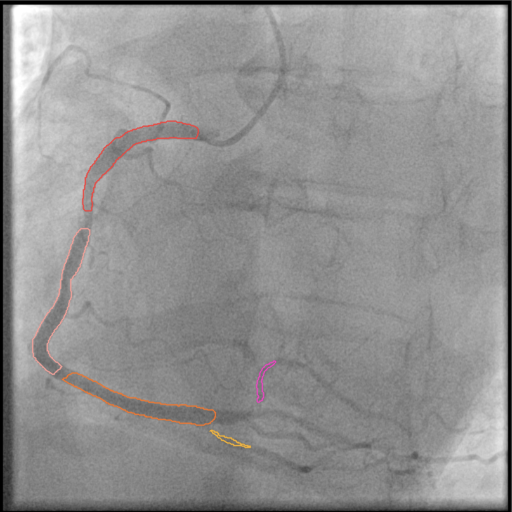} \\

\includegraphics[width = 1.5in]{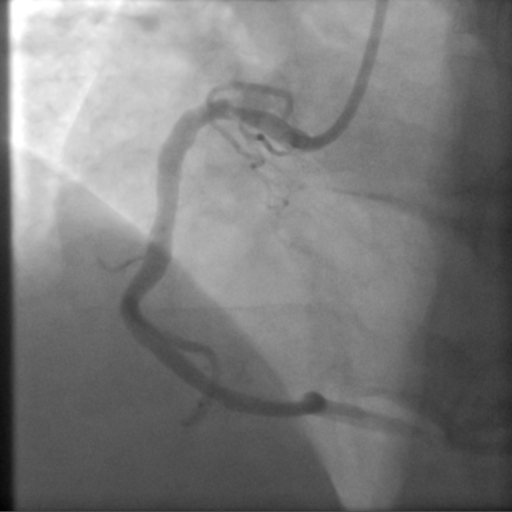} &
\includegraphics[width = 1.5in]{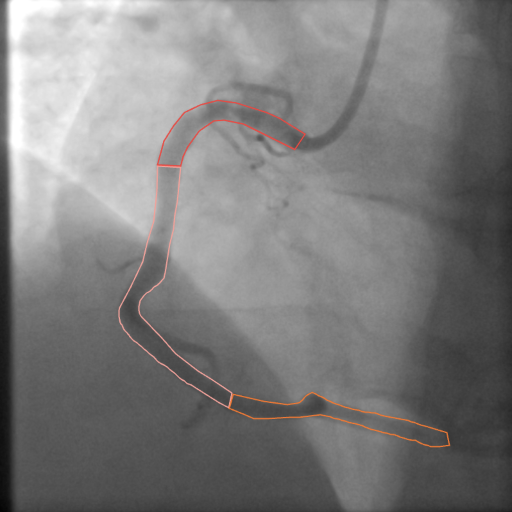} &
\includegraphics[width = 1.5in]{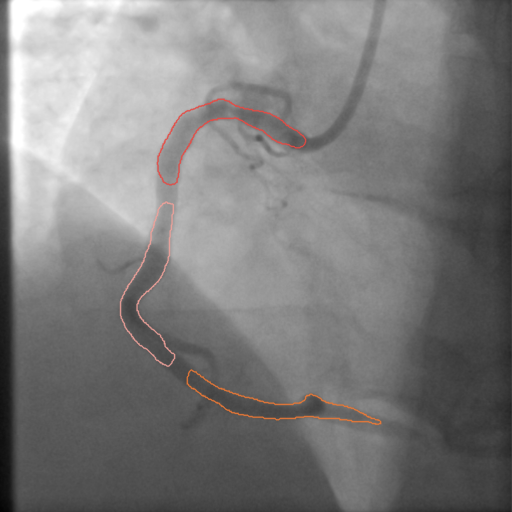} \\

\includegraphics[width = 1.5in]{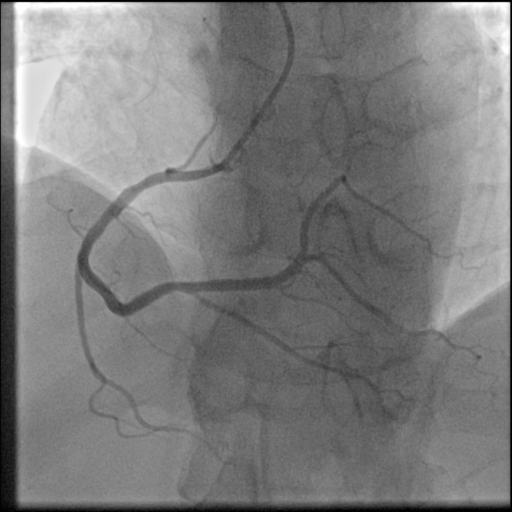} &
\includegraphics[width = 1.5in]{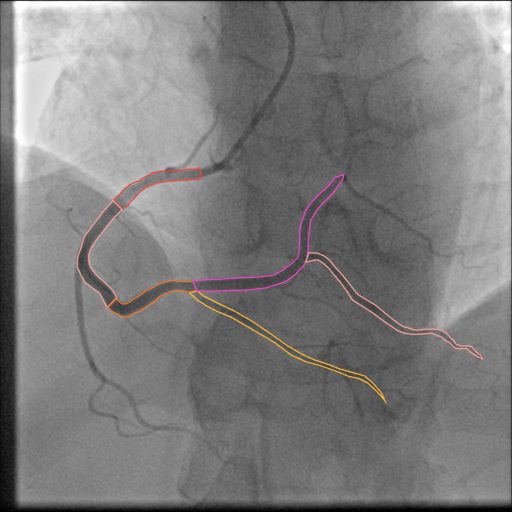} &
\includegraphics[width = 1.5in]{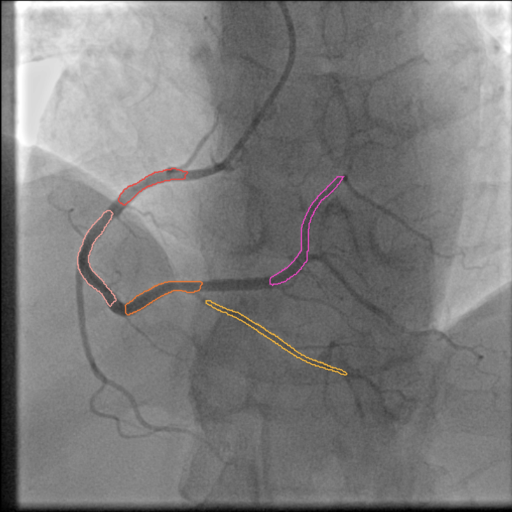} \\

\includegraphics[width = 1.5in]{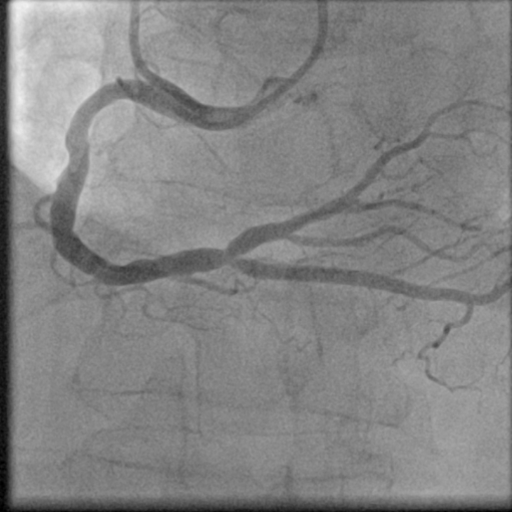} &
\includegraphics[width = 1.5in]{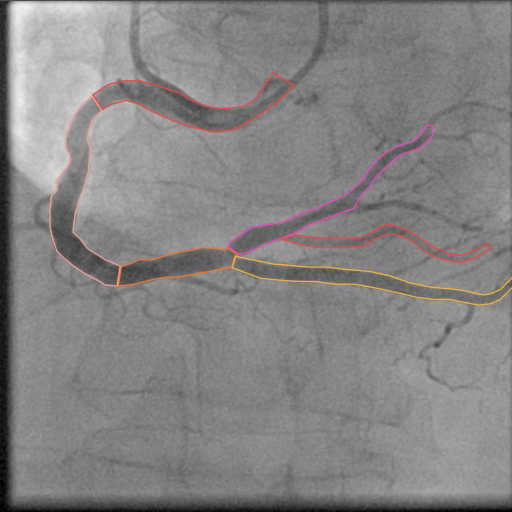} &
\includegraphics[width = 1.5in]{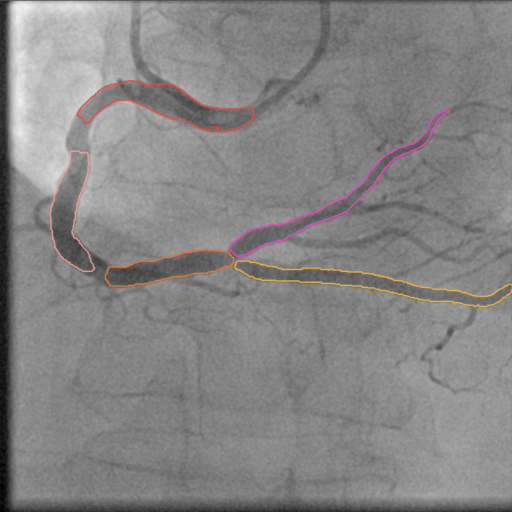} \\

\subfloat[Input]{\includegraphics[width = 1.5in]{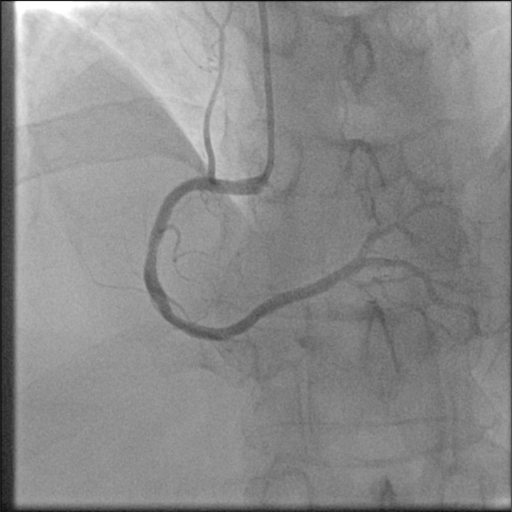}}&
\subfloat[Ground-truth]{\includegraphics[width = 1.5in]{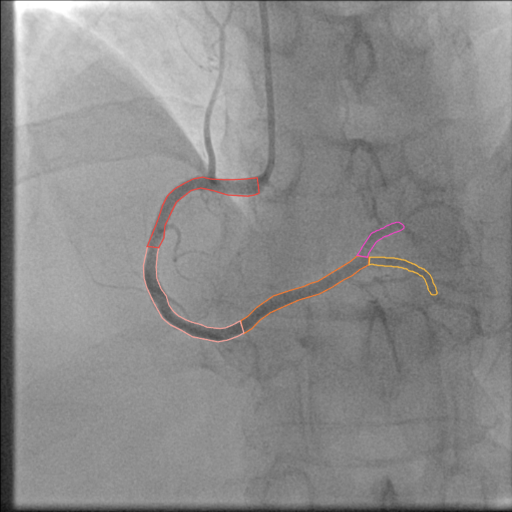}}&
\subfloat[Ours]{\includegraphics[width = 1.5in]{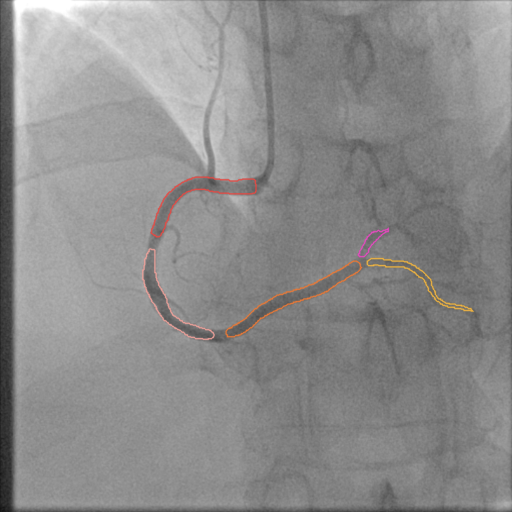}}
\end{tabular}
\caption{Qualitative results of RCA. The model more accurately segments the side branch of RCA than that of LCA. Also, the model tends to predict masks small compared to the ground truth mask. }
\label{fig:qualitative rca}
\end{figure}

\jh{
We evaluated the results of the validation and test dataset. As shown in Table \ref{accuracy of vessel classification}, the vessel classification model achieved an exceptionally high accuracy. Upon inspecting misclassified data, we observed a tendency that our model makes incorrect predictions, when the guide wire was positioned in a location similar to the blood vessel. Such data were not abundant in the training dataset, and due to the resemblance in pixel value and appearance between the wire and the blood vessel, the model often misinterpreted the wire as a blood vessel, leading to classification failures.

\begin{table}[h]
\caption{Accuracy of vessel classification in the validation dataset.}
\label{accuracy of vessel classification}
\centering
\begin{tabular}{|c|c|}
    \hline
    Model & Accuracy \\
    \hline
    YOLOv8m & $0.990$ \\
    \hline
\end{tabular}
\end{table}

In the case of SYNTAX segmentation, the ensemble model had a notable difference of 0.11 points compared to using a single model for predicting the validation data (Table ~\ref{mean f1 of segment}).
This observation underscores the significant advantage of ensemble predictions, where the average predictions from multiple models contribute to more robust detection, ultimately enhancing the accuracy and reliability of the segmentation results.

\begin{table}[h]
\caption{Mean F1 scores of each vessel in the validation dataset.}\label{mean f1 of segment}
\centering
\begin{tabular}{|c|c|c|}

\hline
Vessel type & Model & mean F1 score\\
\hline
RCA & Unet++(resnet50)  &  $0.592$  \\
RCA & Unet++(resnet34)-1 & $0.569$  \\
RCA & Unet++(resnet34)-2 & $0.518$  \\
RCA & Unet++(resnet34)-3 & $0.547$  \\
RCA & Ensemble & \textbf{$0.629$}  \\
\hline
LCA & Unet++(resnet50)  &  $0.386$ \\
LCA & Unet(resnet34) & $0.385$   \\
LCA & Unet++(resnet34) & $0.387$  \\
LCA & Ensemble & \textbf{$0.419$} \\
\hline
\end{tabular}
\end{table}

For LCA, combining predictions from the SYNTAX segmentation with the SYNTAX segment classification model, which has an accuracy of 86\%, as reflected in Table~\ref{accuracy of segment classification}, did not lead to a significant improvement in the final mean F1 score. 
The decision to integrate the SYNTAX segment classification model was driven by the aim of enhancing the classification accuracy of the side branch. Unfortunately, the SYNTAX segmentation model did not generate predictions specifically for the side branch, leading to an outcome where the anticipated performance improvement was not achieved.

\begin{table}[h]
\caption{Accuracy of SYNTAX segment classification in the validation dataset.}
\label{accuracy of segment classification}
\centering
\begin{tabular}{|c|c|}
    \hline
    Model & Validation \\
    \hline
    EfficientNet-b3 & $0.847$\\
    ResNet34 & $0.823$\\
    DesnseNet121 & $0.843$\\
    Ensemble & \textbf{$0.863$}\\
    \hline
\end{tabular}
\end{table}

When comparing the performance across different vessel types, the mean F1 score of RCA outperformed the score of LCA by 0.21 points based on the validation set. As previously mentioned, RCA had a simpler vascular structure and more consistent shooting angles in imaging, making it easier for the model to train and predict. 
In contrast, the segmentation of LCA poses unique challenges due to wide variations in the positions and complex structures of vessels, even within the same class segment. 
Additionally, in certain cases, labels were not accurately assigned to images with less complex vascular structures. These various factors collectively presented significant difficulties in training the model for LCA vessels, resulting in the observed performance gap between RCA and LCA.

In Fig. \ref{fig:qualitative rca} and \ref{fig:qualitative lca}, most of the predicted masks are relatively smaller in size compared to the ground truth masks. 
While predictions are not made in areas without contrast agents, there is a tendency to make less accurate predictions in regions with contrast agents present. 
Interestingly, no overlapping regions between SYNTAX segments were found, but there are noticeable gaps or inaccuracies in delineating the boundaries between individual SYNTAX segments.

To address this problem, we conducted training a model that takes masked images of blood vessels as inputs to predict SYNTAX segmentation. When training the model to predict the SYNTAX segment of LCA using ground truth masks, we observed a relatively higher score of 0.61 on the validation dataset compared to the previous 0.42 score.

Subsequently, we attempted to train a segmentation model for vessel mask prediction by consolidating SYNTAX segment masks into a single representation as the complete vessel mask. 
Despite efforts to optimize, the model's performance yielded a low intersection over union (IoU) of 0.45. This suboptimal outcome was primarily attributed to the absence of masks for the entire vasculature in the training data.
This issue causes disruptions in the model training process. We anticipate that training the segmentation model with data that includes the entire vessel mask will lead to enhanced performance in LCA segmentation.
}

\section{Conclusion}

\hw{
In this paper, we introduced a multi-stage methodology that takes into account the intricate structure of the cardiovascular system for conducting coronary tree segmentation based on the SYNTAX Score in coronary angiography. 
The proposed method initially classifies vessels into LCA and RCA, which possess morphologically distinct characteristics and hence cannot be grouped into identical classes. 
Following this classification, dedicated ensemble models for each LCA and RCA carry out the segmentation task. 
In the final step, considering the inherent complexity of LCA compared to RCA, a refinement model makes an additional round of class predictions for segmented areas.
Our approach not only exhibits potential for application to various future segmentation models but also promises enhanced performance levels. 
Most notably, our approach has showcased exceptional performance and efficacy in the ARCADE-Segmentation Detection Algorithm. \newpage
}

\bibliographystyle{splncs04}
\typeout{}
\bibliography{main}

\end{document}